%% file: main.tex
\documentclass{article}

\input{preamble.tex}
\title{Treatment effect estimation by comparing observed and predicted outcomes: conditions for valid inference and practical illustration}
\author{
Lotta M. Meijerink$^{a, *}$ \and 
Artuur M. Leeuwenberg$^{a}$ \and 
Jungyeon Choi$^{a}$ \and 
Bas B. L. Penning de Vries$^{a}$ \and 
Johannes A. Langendijk$^{b}$ \and 
Judith G.M. van Loon$^{c}$ \and 
Remi A. Nout$^{d}$ \and 
Karel G.M. Moons$^{a}$ \and 
Ewoud Schuit$^{a}$
}

\begin{document}

\maketitle
\begin{center}
\small

$^{a}$Julius Center for Health Sciences and Primary Care, University Medical Center Utrecht, Utrecht University, Universiteitsweg 100, 
3508 GA Utrecht, The Netherlands \\[0.8em]

$^{b}$Department of Radiation Oncology, University Medical Center Groningen,  
Hanzeplein 1, 9713 GZ Groningen, The Netherlands  \\[0.8em]

$^{c}$MAASTRO Clinic, Doctor Tanslaan 12, 6229 ET Maastricht, The Netherlands \\[0.8em]

$^{d}$Department of Radiotherapy, Erasmus MC Cancer Institute, 
Dr Molewaterplein 40, 3015 GD Rotterdam, The Netherlands  \\[0.8em]

$^*$\textit{Corresponding author (l.m.meijerink-20@umcutrecht.nl)}
\end{center}

\maketitle
\newpage

\section*{Abstract}
\subsubsection*{Background}
One non-randomized approach to estimate the average effect of a newly introduced treatment is to compare \textit{observed} outcomes under the new treatment with \textit{predicted} outcomes under the standard treatment. These counterfactual predictions are made using a model developed before the new treatment was introduced, using patient characteristics and individualized treatment information. 
Although the approach has been intuitively applied (e.g., as model-based clinical evaluation in radiotherapy) and can be recognized as a specific case of standardization, a method of virtual controls or a g-method, the theory and conditions required for unbiased treatment effect estimation have not been formally described. The objective of this paper is to formalize the approach and clarify these conditions.
\subsubsection*{Methods}
We formalize the approach within the potential outcomes framework for causal inference. We explain the methodology, its necessary conditions, and approaches for assessing their validity. These conditions are furthermore illustrated through a case study from radiotherapy, estimating the benefit of proton therapy compared to photon therapy on dysphagia in patients with head and neck cancer.
\subsubsection*{Results}
We describe a set of five sufficient conditions, including examples of violations: transportability, ignorability of treatment assignment, consistency, positivity, and correct model specification. While these conditions are largely untestable, we describe how empirical evidence, such as comparing predicted and observed outcomes in related samples, can increase confidence in their plausibility. 
\subsubsection*{Conclusion}
When the prediction model predicts well in relevant (sub)populations, the approach can yield unbiased treatment effect estimates. However, there are many possible sources of bias. Therefore, we recommend systematic consideration of all required conditions, informed by domain expertise, and the use of empirical evidence whenever possible to support their plausibility.

\bigskip
\paragraph{Keywords} model-based clinical evaluation, counterfactual prediction, average treatment effect among the
treated, ATT, virtual controls, transportability

\newpage

\section{Introduction}
New treatments need extensive evaluation to confirm their (cost-)effectiveness and safety. While randomized controlled trials (RCTs) are considered the gold standard for estimating treatment effects \cite{hariton_randomised_2018}, they are not always feasible or ethical, or may not yet be available at the time decisions about the new treatment need to be made \cite{saesen_defining_2023, noauthor_evaluation_2014}. Consequently, treatments are sometimes introduced before trial evidence is available \cite{langendijk_clinical_2018}.  In such cases, non-randomized methods, e.g., regression adjustment or propensity score techniques,  can be used to generate evidence on treatment effectiveness, either while awaiting trial results, or to complement existing evidence. 

An alternative approach focuses on the set of patients who received the newly introduced treatment and applies a model to predict their counterfactual outcomes: \textit{what would their outcome have been had they received the standard treatment instead?} The average treatment effect among the treated (ATT) is estimated as the average difference between observed outcomes under the new treatment and predicted outcomes under the standard treatment. In this approach, the model used for counterfactual predictions is based on historical or external patient data from a setting where the new treatment has not yet been implemented, i.e., where everyone received the standard treatment. The approach can be viewed as a g-method \cite{hernan_causal_2020}, a method of virtual controls \cite{strayhorn2021virtual} or an application of model-based standardization \cite{dahabreh_generalizing_2019, lash_modern_2021}.

While the approach (comparing observed outcomes under the new treatment with predicted outcomes under the standard treatment) seems intuitive and practical, and has been applied in several medical domains including oncology, cardiology, and mental health research \cite{lester-coll_modeling_2019, rwigema_model-based_2019, christianen_swallowing_2016, jia_generation_2014, neal_discriminating_2013, ford_strengths_2009, ketchum_predictive_2010}, it relies on several conditions that are often not explicitly discussed. 

The aim of this paper is to clarify these conditions. We do this by formalizing the approach in the potential outcomes framework and discussing a set of sufficient conditions: what they mean, why we need them, and when they would be violated. 
To illustrate the approach and its conditions more concretely, we discuss them using a realistic case study from radiotherapy (introduced below), where the approach was proposed as model-based clinical evaluation, or a model-based approach \cite{langendijk_clinical_2018, meijer_reduced_2020}, and is used to estimate the effect of new technologies, such as proton therapy or swallowing sparing intensity modulated radiotherapy, in reducing radiation-induced complications \cite{lester-coll_modeling_2019, rwigema_model-based_2019, christianen_swallowing_2016}. 

\section{Case study}\label{sec:casestudy}
Radiation technology is constantly developing, often aiming to reduce the radiation dose to healthy organs surrounding the tumor, and thereby preventing radiation-induced complications, while maintaining the required dose to the tumor itself. 
An example is proton therapy, which allows for more precise radiation delivery - specifically, less dose to surrounding healthy organs - compared to `conventional' photon therapy (such as VMAT: Volumetric Modulated Arc Therapy). Clinicians widely agree that in general, reducing radiation exposure to healthy tissues is beneficial, following the ALARA principle that radiation exposure should be "as low as reasonably achievable". 

As a result, innovative radiotherapy technologies are often adopted relatively quickly, sometimes even without formal evaluation through RCTs \cite{langendijk_clinical_2018}. Furthermore, there are additional barriers to conducting RCTs on new radiation technologies. For example, unlike pharmaceutical companies, medtech firms have limited financial incentives to fund trials, as successful technologies generally benefit all providers rather than just the one investing in the often costly study.
Nonetheless, especially when investment costs are considerable, gathering evidence of the benefits of the new technologies remains important \cite{rwigema_model-based_2019}. 
\subsection{Research aim}
This case study is based on (but a simplified version of) a real-world setting in the Netherlands and focuses on patients with head and neck cancer undergoing radiotherapy. All example calculations are based on synthetic data.
Interest was in the estimation of the benefit of proton therapy in terms of reducing dysphagia (difficulty swallowing) at 6 months after radiotherapy, compared to photon therapy (VMAT), or more specifically, in the average benefit in the patients currently eligible for proton therapy. Here, eligibility was determined by `model-based selection' \cite{langendijk_selection_2013}, which is further explained below.

\subsection{Population} \label{sec:casestudy-data} Patients were treated in a single center, covering two time periods:
\begin{itemize}
\item \textbf{2007-2017 (Pre-introduction sample):} All 750 patients treated during this period received photon-based radiotherapy (VMAT).
\item \textbf{2018-2019 (Post-introduction sample):} During this time, proton therapy became available as a treatment option, but due to its higher costs and limited capacity, was only reimbursed for patients who were expected to benefit significantly from it. Not all patients are expected to benefit equally from proton therapy. For example, in some cases, the tumor is located such that conventional photon therapy can target it effectively, without causing much damage to nearby healthy tissues. In such situations, the added value of the more expensive proton therapy is minimal. However, for other patients, proton therapy can substantially reduce the radiation dose delivered to surrounding healthy organs and thereby reduce radiation-induced toxicity.

A model-based selection approach was introduced to determine which patients would be eligible to receive proton therapy.\cite{langendijk_selection_2013} In this approach, for each patient, two radiation plans were created: one using photon therapy and one using proton therapy. From these plans, mean planned doses to tissues surrounding the tumor (the superior, middle, and inferior pharyngeal constrictor muscles and the oral cavity) were extracted. These dose metrics, along with two patient-specific characteristics, baseline dysphagia status and tumor location, served as input for an existing logistic regression model to predict the risk of dysphagia at 6 months after radiotherapy. The difference in predicted dysphagia risk between the photon and proton plans was used as an estimate of the expected benefit of proton therapy. Patients were selected for proton therapy if this expected benefit exceeded 10\%.
To summarize, selection for proton therapy was based directly on the planned dose parameters from both treatment plans and the two baseline characteristics. Using this procedure, 93 patients were selected for and treated with proton therapy, while 207 received photon therapy during this period.

\end{itemize}
For all patients, data were collected on baseline characteristics, the four photon plan dose parameters, and the grade of dysphagia after 6 months (the outcome). For patients treated after the introduction of proton therapy, additionally, the same four dose parameters were available from the proton plan. We refer to Appendix A.1 for a synthetic example of what the data of this case study could look like, and Appendix B for the description of how these synthetic data were generated.

\section{Methodology}
This section describes the methodology in general, and in the context of the case study. We describe the estimand, how this is estimated,  the sufficient conditions, and finally, ways to gain supportive evidence of the validity of the approach using model validation.

\subsection{Causal estimand}\label{sec:estimand}
The approach is used to compare two treatments:
\begin{itemize}
    \item \( T = 0 \): the standard (control) treatment 
    \item \( T = 1 \): the (new) target treatment under evaluation
\end{itemize}
Let \( Y \in \{0, 1\} \) be a binary outcome, where \( Y = 1 \) indicates the presence or occurrence of a specific health state at the predefined relevant outcome timing. Our causal question is:
\begin{quote}
\textit{To what extent does the target treatment \( T=1 \) reduce the risk of the outcome \( Y=1 \), compared to standard treatment \( T=0 \), in the population currently treated with \( T=1 \)?}
\end{quote}
This corresponds to the average treatment effect among the treated (ATT).  
To formalize this question, we define it using the potential outcomes framework \cite{rubin_causal_2005}. We let \( Y(t) \) denote the potential outcome of an individual under treatment \( T = t \), i.e., the outcome that would be observed if the individual receives treatment $t$. Then, the ATT is defined as:
\begin{equation*}
    ATT 
    = \mathbb{E}_{post}[Y(1) - Y(0) \mid T = 1]
\end{equation*}
i.e., the average difference between the two potential outcomes, in the patients treated with the target therapy. Here, we use the subscript $post$ to denote the `post-introduction' setting, i.e. the setting where the target treatment is already introduced. This setting will be different from the setting that provides us data to develop the model (e.g. because they come from a different hospital or moment in time), which we will refer to as `pre-introduction'. 

Alternative effect measures (e.g., odds ratio or risk ratio) can be defined analogously.

\paragraph*{Case study}
In the case study, the standard treatment ($T = 0$) is photon therapy, and the target treatment ($T=1$) is proton therapy. The outcome $Y$ is dysphagia (grade 2 or higher) at 6 months after radiotherapy. A potential outcome $Y(0)$ for an individual represents whether they would experience dysphagia if they received photon therapy. The population of interest is the population currently treated with protons. We care about the treatment effect in the population currently treated with protons because these are the patients in whom we expected a benefit of proton therapy. The causal contrast of interest is a risk difference, i.e., the expected difference in the proportion of patients with dysphagia if everyone were treated with photons versus if everyone were treated with protons (ATT).

\subsection{Estimation}
\begin{figure}[H]
    \centering
    \includegraphics[width=\linewidth]{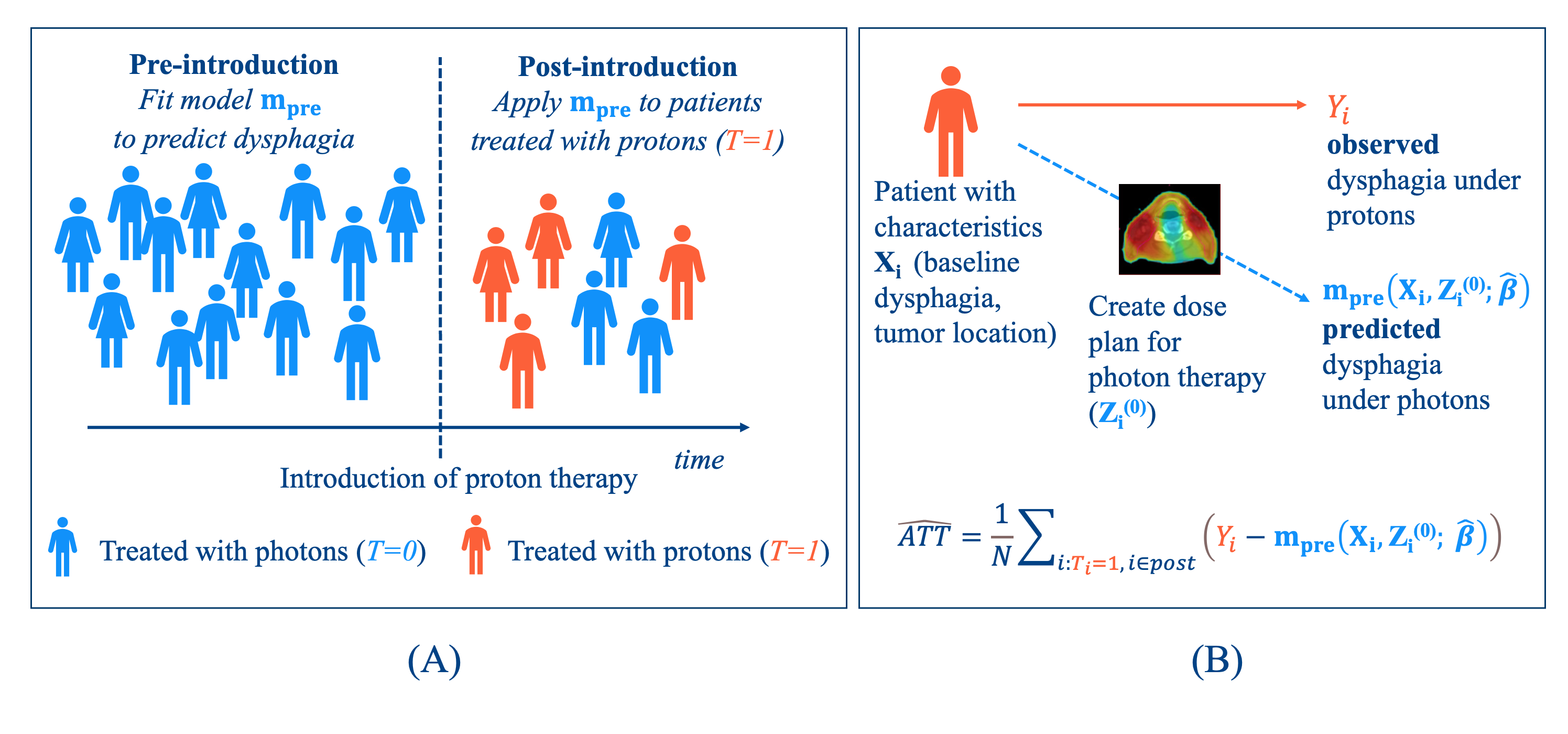}
    \caption{Illustration of the approach as applied in the case study. (A) The patients that were treated before introduction of proton therapy were used to develop the model $\mathbf{m}_{pre}$. These patients were all treated with photons (represented in blue). After introduction of proton therapy, only a subset was treated with protons (represented in orange).  (B) On this proton-treated subset of individuals, the model is applied to make counterfactual predictions using the previously developed model $\mathbf{m}_{pre}$. To make predictions, the model uses four variables from their individual photon dose plan $\mathbf{Z}^{(0)}$ and their other patient characteristics $\mathbf{X}$ (baseline dysphagia and tumor location). The average difference between predicted and observed outcomes is taken as an estimate of the average treatment effect among the treated (ATT). }
    \label{fig:illustration-mbe}
\end{figure}

The estimation procedure in context of the case study is shown in Figure \ref{fig:illustration-mbe}. To estimate the ATT, we make use of a model family $\mathbf{m}_{\text{pre}}(\mathbf{X},\mathbf{Z}^{(0)}, \boldsymbol{\beta})$.  parametrised by $\boldsymbol{\beta}$ (e.g. a logistic regression with a set of coefficients). We fit the model in a historical (or external) cohort where all patients received the standard treatment $T = 0$, giving us fitted parameters $\boldsymbol{\hat{\beta}}$.  This fitted model estimates the risk of the outcome \( Y = 1 \) under treatment $T=0$ given:
\begin{itemize}
    \item \textbf{Patient characteristics} \( \mathbf{X} \), e.g., age, comorbidities
    \item \textbf{Treatment plan variables \( \mathbf{Z}^{(0)} \) of the standard treatment $T=0$}, such as the planned chemotherapy frequency in cancer treatment, or information about the insulin dosing strategy in diabetes management. In case the standard treatment of interest can only have one single operationalization (e.g., a pill with fixed dose), there will be no need to include treatment plan variables. 
\end{itemize}
Let \( i = 1, \ldots, N \) index the sample of patients from the post-introduction population, that are currently treated with \( T = 1 \). Then:
\begin{equation*}
    \widehat{ATT} = \frac{1}{N} \sum_{i=1}^{N} \left( Y_i - \mathbf{m}_{pre}(\mathbf{X}_i, \mathbf{Z}^{(0)}_i ; \boldsymbol{\hat{\beta}}) \right)
\end{equation*}
In words, this estimate is the average difference between observed outcomes and predicted counterfactuals under $T=0$, in the sample of the post-introduction population that was treated with $T=1$.

\paragraph*{Case study}
In the case study, the model family $\mathbf{m}_{pre}$ is logistic regression, with model parameters (coefficients) $\boldsymbol{\hat{\beta}}$ fitted on the 750 individuals treated with photon therapy between 2007 and 2017.
As patient characteristics $X$, this model incorporates the grade of dysphagia at baseline (i.e., before starting radiotherapy) and tumor location. As treatment plan characteristics $\mathbf{Z}^{(0)}$ four variables from the photon radiation plan are used: the mean planned dose to the superior pharyngeal constrictor muscle (PCM), middle PCM, and inferior PCM. The ATT is calculated using the 93 individuals treated with protons in the target sample (2018-2019). For these individuals, although they were treated with protons, corresponding photon plans are also available as these were used selecting patients for proton therapy (see section 2.3). This allows us to compare their observed dysphagia outcome $Y_i$ with the counterfactual risk of dysphagia predicted by the fitted logistic regression model, using their photon plan and patient characteristics: $\mathbf{m}_{pre}(\mathbf{X}_i, \mathbf{Z}^{(0)}_i ; \boldsymbol{\hat{\beta}})$. In Appendix A.4 we show an example calculation in R, using bootstrapping to obtain confidence intervals around the estimated ATT. In this example, the average of observed outcomes under proton therapy was 0.25, whereas the average of predicted counterfactuals under photon therapy was 0.47, resulting in a treatment effect estimate of -0.22, with a 95\% CI of (-0.31, -0.13). This positive ATT is expected, as patients were selected for proton therapy based on their expected benefit.

\subsection{Conditions for validity}
As illustrated above, it is relatively easy to estimate a treatment effect using the approach above, but under what conditions does it provide a valid answer to the causal question?
We describe a set of conditions that ensure validity of the estimation procedure. Weaker conditions could suffice to obtain valid estimates; however, we present the current set of conditions because they offer an intuitive and useful framework for understanding the problem. For more details, including a derivation from causal estimand to the estimator, that shows why these conditions are sufficient, please refer to Appendix C. 
\subsubsection{Condition 1: Transportability} \label{sec:transportability}
This condition requires that the risk of outcome $Y$ under treatment $T=0$ \textit{given covariates} \textit{in the model} are the same in the pre- and post-introduction populations: 
\[P_{pre}(Y(0)=1 \mid \mathbf{X}, \mathbf{Z}^{(0)}) = P_{post}(Y(0)=1 \mid \mathbf{X}, \mathbf{Z}^{(0)})\]
This holds if any variable influencing outcome $Y(0)$ has the same distribution across populations (conditional on covariates) or is correctly incorporated in the model. In general, the relationship between covariates and the outcome also must stay consistent between populations. The condition only applies to $Y(0)$, the outcomes under treatment $T=0$, as only these are predicted using the pre-introduction model. 

\paragraph*{Why?} This condition allows us to apply a model developed in a sample from a historical or external population to make valid counterfactual predictions in the post-introduction population.

\paragraph*{Examples of violations} 
\begin{itemize}
    \item If smoking status is not included in the model but does increase the outcome risk \textit{and} there is a higher smoking prevalence in the pre-introduction population, compared to post-introduction. This would cause the model to overestimate risk when applied to the post-introduction population. 
    \item If there are concurrent changes in treatment protocols (e.g., different opioid prescribing protocols between hospitals or over time) that influence the outcome but aren't included as predictors in the model.
    \item If there are differences in dose optimization strategies between populations, causing differences in the radiation dose to the larynx, and this dose does affect the outcome, but is not incorporated in the model.  

\end{itemize}

\paragraph*{Case study} As described in section \ref{sec:casestudy}, the case study considers a situation where pre- and postintroduction samples came from the same center, so potential differences would primarily have arisen from temporal changes. There were no notable changes regarding concurrent treatment strategies and changes in patient characteristics are small, as only a relatively short time span is considered. A plausible violation would be that over time, dose optimization strategies improved, leading to reduced doses in nearly all organs, including those not explicitly included in the model. 
While the organs most relevant to the outcome of interest (dysphagia) are represented in the model, these temporal changes may still cause an overestimation of predicted outcomes under photon therapy. This would result in an overestimation of the treatment effect, meaning we might detect a larger benefit of protons than truly exists, due to the improvement of the dose planning under the standard treatment, i.e., photon therapy.

\subsubsection{Condition 2: Ignorability of treatment assignment}  \label{sec:ignorability}
This condition requires that within the post-introduction population, the potential outcome that would be observed under standard treatment is independent of the treatment that the patient was actually treated with, given covariates: 
\[Y(0) \independent T \mid \mathbf{X}, \mathbf{Z}^{(0)}\]

\paragraph*{Why?} This condition allows us to apply a model developed in a sample from a historical or external population, where \textit{everyone} received the standard treatment, to make valid counterfactual predictions in a non-random \textit{subset} of the post-introduction population: the individuals that were selected to be treated with the target treatment. If selection was random, this condition is always satisfied.

\paragraph*{Examples of violations} 
\begin{itemize}
    \item If the target treatment is only available to patients with certain socioeconomic characteristics (e.g., only wealthy patients due to insurance limitations) and these characteristics influence the outcome independently of the covariates in our model.
    \item If clinicians assign treatment based on patient-reported quality of life, that is not included in the model, but that does predict the outcome even after accounting for the variables that \textit{are }included in the model.
\end{itemize}

\paragraph*{Case study} Treatment selection for protons followed a deterministic process based on baseline dysphagia, tumor location, planned dose under photons, and planned dose under protons, as described in section 2.3. If the model would have included all these variables used for selection, the condition would be satisfied. However, the model does not include the planned dose variables under protons, as proton plans were not available at the time model development data were collected. 
The question becomes: after conditioning on baseline dysphagia, tumor location, and planned photon dose variables, does knowledge of planned proton dose provide additional information about expected outcomes under photon therapy? Said differently, are there unmeasured variables that correlate with both the potential reduction in dose (from photons to protons) and dysphagia risk? Plausible violations are:
\begin{itemize}
\item Tumor location, which influences both the possible reduction in dose, as well as the risk of dysphagia. Tumor location is included in the model, although it may not be fully accounted for as it is only incorporated as broad categories. 
\item Patients with a higher N-stage often receive higher photon doses and may allow for greater potential dose reductions using protons. At the same time, a higher N-stage may be associated with a higher risk of dysphagia. Because N-stage is not included in the model, the predicted outcomes under photon therapy may be slightly underestimated in the proton-treated group, which could lead to an underestimation of the benefit of proton therapy.
\end{itemize}
To decide whether ignorability of treatment assignment holds, it may help to draw out assumptions about relevant causal relations. An example graphical representation of the case study is shown in Appendix D.

\subsubsection{Condition 3: Consistency} \label{sec:consistency}
This condition requires that for each patient receiving a treatment \(T = t\), their observed outcome (\(Y_i\)) should match their potential outcome under that same treatment:
\[
Y_i = Y_i(t) \quad \text{if } T_i = t \quad \text{ for $t = 0,1$}
\]
This reflects that treatment has a single well-defined meaning, i.e. all realistic ways of delivering a treatment are equivalent in terms of how they affect patient outcomes. In other words, once it is decided that a patient receives treatment \(T = t\), \textit{how} it is delivered (the specific equipment, clinician, or workflow used) should not change the patient's outcome. 

\paragraph*{Why?} Consistency allows us to connect what we observe in the data to the potential outcomes used in causal reasoning, such that we can interpret the estimated effect as the effect of treatment. Lack of consistency would result in an ambiguous treatment effect estimate, as you are effectively comparing different treatments under the same label.

\paragraph*{Examples of violations}
\begin{itemize}
    \item If a patient is prescribed the same medication (e.g., insulin for diabetes), but the delivery method (e.g., insulin pump vs. manual injection) is not part of the treatment definition and would have led to different outcomes.
    \item If the clinical outcome of a patient depends strongly on the clinician's skill or interpretation, even for the same treatment (e.g. a specific type of surgery, or radiotherapy planning).
    \item If an older and a newer radiation machine deliver the same therapy (e.g., photon radiotherapy) but approximate the planned dose with different levels of accuracy that would lead to different outcomes. 
\end{itemize}

\paragraph*{Case study} We assume that all meaningful variations of photon therapy are reflected by differences in dose distributions, which are included as covariates in the model. Some variations may still exist, even for the same planned dose distributions, e.g. due to variations in fractionation, but we believe it is a reasonable assumption that those variations would not result in a different grade of dysphagia at 6 months.

\subsubsection{Condition 4: Positivity} \label{sec:positivity}
This condition requires that all combinations of values of covariates in the model (both patient characteristics \(\mathbf{X}\) and treatment plan variables under the standard therapy \(\mathbf{Z}^{(0)})\), that could occur among the individuals in the post-introduction population that received the target treatment, could also occur in the pre-introduction population: 
\[
\text{Supp}\left[ (\mathbf{X}, \mathbf{Z}^{(0)}) \mid T=1 \right]_{\text{post}} \subseteq \text{Supp}\left[ (\mathbf{X}, \mathbf{Z}^{(0)}) \right]_{\text{pre}}
\] 
Here, the support of a distribution refers to the set of values or combinations that the variables can take with non-zero probability. This is a somewhat theoretical condition: it does not require that all combinations actually occur in the data, but that there are no structural reasons preventing them from occurring.

Even if the positivity condition holds in theory, it may happen that some combinations are not observed in the available data simply due to chance, especially when the sample size is limited or certain subgroups are rare. This is known as a stochastic positivity violation \cite{petersen_diagnosing_2012, zivich_positivity_2022}. Although such stochastic violations do not invalidate the theoretical identifiability of the treatment effect (meaning that, with infinite data, the effect could still be estimated), they can cause similar practical challenges and should also be considered.

\paragraph*{Why?} To make reliable counterfactual predictions for patients in the target treatment group (i.e., those who received $T=1$), their characteristics should be adequately represented in the population used to develop the model (the pre-introduction group). If certain combinations of patient features or treatment variables never appear in the pre-introduction population, the model needs to rely on extrapolation, essentially "guessing" outcomes for patients unlike anyone seen before. Depending on further parametric assumptions (e.g., linearity), this can lead to biased predictions or even make it impossible to make predictions, for example, if a patient has a covariate value that was completely absent during model development.

\paragraph*{Examples of violations} 
\begin{itemize}
    \item If the pre-introduction population comes from a hospital that does not treat patients with oropharyngeal tumors, but the target treatment population includes such patients, and tumor location is included as a covariate in the model. In this case, the model cannot be used to make predictions in these patients.
    \item If the pre-introduction population is strictly adult (18+), but the target treatment population also contains teenagers, and age is used as a covariate in the model. Although this is an example of the positivity violation, it may not \textit{necessarily} prevent making valid counterfactual predictions for the younger patients if the model used is capable of extrapolating to these younger ages (e.g., it can generate predictions for 16-year-olds). The validity of such extrapolations, however, depends heavily on whether the parametric assumptions underlying the model's extrapolation are appropriate.
\end{itemize}

\paragraph*{Case study} We expect this condition to hold because the pre-introduction sample comes from the same hospital as the sample of proton-treated individuals, differing only in time period. 
To decide if positivity is likely to hold, it is possible to examine the univariable distributions of covariates (see Appendix A.2). Although looking at each variable independently does not tell the whole story, together with knowledge that the data comes from the same hospital, it can provide sufficient confidence that there are no issues with positivity.

\subsubsection{Condition 5: Correct model specification} \label{sec:modelspec}
We require the outcome model $\mathbf{m}_{pre}$ to be correctly specified (in terms of functional form) or sufficiently flexible to capture the conditional probability $P_{pre}(Y=1\mid\mathbf{X}, \mathbf{Z}^{(0)})$, i.e., that there are model parameters $\boldsymbol{\beta_0}$ such that $\mathbf{m}_{pre}(\mathbf{X}, \mathbf{Z}^{(0)};\boldsymbol{\beta_0}) = P_{pre}(Y=1\mid\mathbf{X}, \mathbf{Z}^{(0)})$. Note that this condition concerns the relationship between included covariates and the outcome, not whether additional covariates would improve the model.

\paragraph*{Why?} Since we use a statistical model to approximate conditional probabilities, misspecification of this model will lead to biased estimates of these probabilities and consequently biased ATT estimates.

\paragraph*{Examples of violations} 
\begin{itemize}
    \item If the age relates to the outcome in a non-linear way but is modeled with a linear term.
    \item If there is an interaction between tumor location and sex (e.g., a particular tumor location increases risk more in men) but the model is not flexible enough to capture this interaction. For example, it includes tumor location and sex as separate variables in a logistic regression without their interaction.
\end{itemize}

\paragraph*{Case study} A relatively simple parametric model is used, and while its functional form was chosen based on the absence of strong evidence for model misspecification in the pre-introduction data (i.e., no clear non-linearity in the association between dose parameters and the outcome was observed), it is still likely to be somewhat imperfectly specified.
Since we are interested in the average predicted risk across the entire proton-treated population, some errors may average out, potentially limiting bias. Nonetheless, sensitivity analyses examining how different modeling choices affect the resulting ATT estimate would be advisable. We provide an example of such a sensitivity analysis in Appendix A.8.

\subsection{Supportive evidence for validity}
To estimate the average treatment effect among the treated, we make predictions about potential outcomes that are not observed and never will be. For example, in the case study, for patients who received proton therapy, we estimated what would have happened if they had instead received photon therapy, which we never observe. This means that we can never directly validate our estimation strategy. However, in some settings, auxiliary data can be used for indirect validation. 

\subsubsection{Calibration-in-the-large}
If some patients in the post-introduction population received the standard treatment, it is possible to assess how well the model predicts outcomes under the standard treatment in this sample. One option is to use the same estimation procedure as before to estimate the ATT, but now in these individuals who actually received the standard treatment, i.e., estimate the average difference between observed and predicted outcomes under the standard treatment:
$
  \frac{1}{N} \sum_{i: T_i = 0\color{black}, i \in \text{post}} \left( Y_i - \mathbf{m}_{\text{pre}}(\mathbf{X}_i, \mathbf{Z}^{(0)}_i; \boldsymbol{\hat{\beta}}) \right)
$.
In prediction model literature, where validating models in an external data source is common practice, this is known as mean calibration or calibration-in-the-large (CITL) \cite{van_calster_calibration_2019}. This validation strategy also relates to the use of negative control populations \cite{piccininni_using_2024}, where the estimation procedure (in our case, the estimation of the ATT) is repeated on a group of individuals assumed to be similar to the population of interest, except that we expect the treatment effect to be zero.
\\
The CITL (average difference between observed and predicted outcomes) should be interpreted as follows:
\begin{itemize}
\item Difference close to 0: If all assumptions hold, we expect the CITL to be close to zero. However, the converse is not necessarily true: good calibration in this group does not guarantee unbiased ATT estimates, since those are based on predictions for a different subgroup (the ones treated with protons), and this subgroup could systematically differ. Also, limited sample sizes will often limit the ability to draw strong conclusions. A close agreement between observed and predicted outcomes therefore only provides supportive evidence for the robustness of the estimation strategy.
\item Systematic difference: If we do observe a systematic difference (e.g., consistent over- or underestimation), this would reflect violations of one or more conditions. One possibility is to assume that this difference is a general shift in outcome risk (e.g., due to setting or time) that also holds in the patients treated with the target treatment and use it to adjust the treatment effect estimates. When using logistic regression, an option to implement this is to update the model intercept based on the patients receiving the standard treatment.
\end{itemize}
Besides looking at patients from the post-introduction population who were treated with the standard treatment, more opportunities for obtaining supportive evidence for validity may exist. An example is given below in context of the case study.
\paragraph*{Case study}
In the case study, the CITL in the photon-treated subpopulation treated after the introduction of proton therapy was 0.0 (95\% CI -0.05 to 0.05), providing no evidence that the conditions were violated. As an additional validation opportunity, we can take advantage of the fact that both proton and photon therapy are forms of radiotherapy that mainly differ in the delivered dose. If we assume that the relationship between dose and outcome learned from photon therapy also applies to proton therapy, we can assess model performance directly in the treated group (proton patients), even though the model was never trained to predict outcomes under proton therapy. The CITL in the proton-treated population was -0.05 (95\% CI -0.16 to 0.05). This means that the predicted risk under proton therapy slightly overestimates the observed outcomes: on average, the predicted risk is approximately 5\% higher than observed. 
If we are willing to assume that this \textit{overestimation} is not due to differences between radiotherapy types, but would also apply to the same individuals had they received photon therapy, this would imply an \textit{overestimation} of the treatment effect. However, this interpretation relies on additional strong assumptions, and uncertainty remains high due to the limited data. Both types of validation are further illustrated in the case study in Appendix A.6 and Appendix A.7.

\subsubsection{Other model performance metrics}
Other methods for assessing calibration, such as calibration curves, can also be informative, provided the sample size is sufficient: if all underlying assumptions hold, we expect the model to be well-calibrated overall. In contrast, discrimination metrics such as the area under the ROC curve (AUROC) do, by themselves, not necessarily offer insight into the validity of assumptions of the validity of ATT estimation using the model. For example, a model may show poor discrimination (e.g., an AUROC near 0.5) due to a restricted or homogeneous case-mix, even if the assumptions are met.

\section{Discussion} 
We have described the conditions under which a model fitted on pre-introduction patient data can be used to estimate the average treatment effect among the treated (ATT), by predicting counterfactual outcomes, i.e., what would have happened had the treated individuals received the standard treatment instead. 

Although these conditions are largely untestable and easily violated, it is important to acknowledge that every study design relies on assumptions. In certain contexts, the conditions required for this approach may be more defendable than alternatives. 
One intuitive study design in situations where data pre- and post-introduction of a treatment is available, may be a simple comparison of event rates before and after the introduction of a new treatment. However, this approach assumes that all changes in outcomes can indeed be attributed to the introduction of the therapy. In contrast, using a model for the outcome allows for adjustment for outcome shifts due to changes in covariate distributions (e.g., population aging) or changes in treatment plans for the standard treatment (e.g., improved dose planning), or for measured differences in population.
Another option could be to compare patients receiving the target versus the standard treatment during the post-introduction period, e.g., using regression adjustment or weighting methods. This requires overlap in all relevant patient characteristics across groups, which may be difficult to assume, e.g., when treatment assignment is based on a deterministic process like in the case study. By using a model trained on pre-introduction data, we can still indirectly compare with patients like the ones currently receiving the new treatment, but who received the standard treatment.

The approach has other strengths. First, in principle, only the model needs to be shared or published; access to original patient-level data from the pre-introduction population is not required. However, estimating confidence intervals of the final ATT estimate may be difficult, since the sampling variance in model development is often not obtainable. Second, the approach is flexible regarding the effect measure (e.g. risk ratio, odds ratio). Finally, a wide range of modeling strategies can be used, including flexible methods like neural networks when large pre-introduction datasets are available.

It depends on the context whether the approach is appropriate, and what level of evidence is needed to justify the untestable assumptions. Although we illustrated some ways to gather supportive evidence for the required assumptions, further work could explore the use of informed bias analyses and the derivation of plausible bounds on the causal effect \cite{lash_bias_2021, nguyen_sensitivity_2017, dahabreh_sensitivity_2023}. These approaches involve the formal, quantitative assessment of how realistic systematic errors may affect an estimate, by explicitly modeling these errors and propagating their impact to obtain bias-adjusted estimates or bounds.
 This would help communicating not only sampling uncertainty but also uncertainty in the assumptions themselves, thereby improving transparency and practical relevance. \\
Furthermore, even when the approach is considered appropriate, there is still room for methodological refinement. Future research could examine strategies to improve model transportability, for example, using anchor regression or related methods \cite{shen_causality-oriented_2023, rothenhausler_anchor_2021}, or develop doubly robust variants of the estimator \cite{bang_doubly_2005, van_der_laan_targeted_2011,robins_estimation_1994}. This could potentially lead to more robust and efficient estimation of the ATT.

Our work makes several contributions. First, although this strategy has been proposed and used as model-based clinical evaluation within radiotherapy research \cite{lester-coll_modeling_2019, rwigema_model-based_2019, christianen_swallowing_2016, langendijk_clinical_2018, meijer_reduced_2020}, we are the first to formalize this specific strategy within the potential outcomes framework for causal inference, and to clarify a set of sufficient conditions.

Second, while similar strategies have been extensively discussed in causal inference literature under, for example, historical controls \cite{pocock_combination_1976}, model-based standardization or g-methods, \cite{hernan_causal_2020,lash_modern_2021}, the specific strategy described here has not, to our knowledge, been formally described. What distinguishes the present strategy is that it explicitly relies on a model estimated in a pre-introduction population, thereby introducing a transportability aspect that is not commonly part of standardization or g-computation approaches for estimating treatment effects. Furthermore, the strategy additionally allows for the incorporation of individualized treatment planning information available prior to treatment.
 
By presenting this relatively specific approach within a formal causal inference framework, while illustrating it through a case study from a field that uses it in practice, we aim to contribute both to the conceptual understanding and practical applicability of the approach.  
  Finally, by complementing theory and examples with a detailed applied example in R we provide additional practical guidance.

There are several limitations to acknowledge. First, we primarily focused on the theoretical conditions required for identification and did not focus on how violations (e.g. model misspecification) translate into bias in practice. Recent work by Choi et al. (2026)
provides further insight into this through simulation studies, specifically considering the situation where variables related to treatment selection are missing in the model \cite{choi_missing2026}. 
Second, we have focused specifically on the ATT in a setting with binary outcomes. Although many of the ideas directly translate to other outcome types (e.g. time-to-event or continuous outcomes) and estimands, further work is needed to fully adapt the theory to these cases.
Third, although we aimed to maintain practical applicability, our primary focus was causal identifiability. To concentrate on conditions for causal identifiability, the case study, while inspired by the Dutch context, represented a simplified version of reality. We acknowledge that real-world analyses often involve additional complexities, such as missing data, competing risks, measurement error, or limited sample sizes, which may introduce further challenges. 

To conclude, we have discussed the theory behind an estimation approach for the ATT: using a model trained on pre-introduction data, to predict counterfactual outcomes in patients treated with the target treatment. While this approach can offer practical and conceptual advantages, the results could be biased if certain conditions are not met. Therefore, when using the approach, we recommend a systematic evaluation of all these required conditions in collaboration with clinical and methodological experts. Where possible, empirical evidence should be gathered to determine whether the conditions are likely to hold, and if not, what impact this has on the conclusions. These practices will improve the credibility and transparency of the treatment effect estimates and ultimately support more informed clinical decisions.

\section*{Declarations}

\paragraph{Ethics approval and consent to participate} No patient data was used for the manuscript, except to generate synthetic data for the case study in R in the appendix. The data generating mechanism for the case study was based on data described in \cite{van2021comprehensive}, where the requirement of informed consent was waived by the ethics committee.
\paragraph{Consent for publication} Not applicable
\paragraph{Availability of data and materials} Code to generate synthetic data as used in the appendix is provided as supplementary material. 
\paragraph{Competing interests} 
All authors have completed the ICMJE uniform disclosure form and declare: support from ZonMW for the submitted work. RN reports research grants and honoraria from Elekta, Varian, Accuray, Sensius, dr Sennewald, MSD, and GSK, paid to their institution; JAL reports research grants from the European Union and the Dutch Cancer Society, consulting fees and honoraria paid to UMCG Research BV by IBA, is chair of the safety monitoring board of the UPGRADE trial (UMC Nijmegen), member of scientific advisory committees for IBA and RaySearch, and reports departmental research collaborations with IBA, RaySearch, Elekta, Mirada, and Siemens; Other
authors declare no financial relationships with any organisations that might have an interest in the submitted work in the previous three years; no other relationships or activities that could appear to have influenced the submitted work.
\paragraph{Funding}
This project has received funding from ZonMw HTA Methodology grant number 10580012210025 (WhyMBA). This dissemination reflects only the authors’ views and the Commission is not responsible for any use that may be made of the information it contains.
\paragraph{Authors' contributions}
ES and AML conceptualized the initial study. LMM drafted the manuscript and created the figures and code under supervision
of ES, AML and KGM. All other authors critically reviewed the manuscript and contributed to revisions. ES is the guarantor of
this work.
\paragraph{Acknowledgements}
Not applicable

\printbibliography

\begin{singlespace}

\newpage

\end{singlespace}

\end{document}

%% file: preamble.tex
\usepackage{graphicx} 
\usepackage{amsmath}
\usepackage{url}
\usepackage{amssymb} 
\usepackage{mathtools}
\usepackage{todonotes}
\usepackage{pgf}
\usepackage{tikz}
\usepackage{tabularx}
\usepackage[mathlines]{lineno}
\usepackage{booktabs}
\usepackage[margin=1in,footskip=0.25in]{geometry}
\usetikzlibrary{arrows, shapes.arrows, shapes.geometric, shapes.multipart,
decorations.pathmorphing, positioning, swigs}
\usepackage{inputenc}
\usepackage{tcolorbox}
\usepackage{float}
\usepackage[
backend=biber,
style=numeric-comp,
sorting=none,
]{biblatex}

\usepackage{setspace}

\usepackage{xr}

\newcommand\independent{\protect\mathpalette{\protect\independenT}{\perp}}
\def\independenT#1#2{\mathrel{\rlap{$#1#2$}\mkern2mu{#1#2}}}